Hailee Carter

Georgetown University Walsh School of Foreign Service, Washington, D.C., USA

Cognitive Sovereignty and the Neurosecurity Governance Gap: Evidence from Singapore

**ABSTRACT**

As brain–computer interfaces (BCIs) transition from experimental medical systems to consumer and military-adjacent technologies, they introduce a novel security domain in which the human nervous system becomes a networked, contestable substrate. Existing frameworks for cybersecurity, biomedical safety, and data protection were not designed to address adversarial threats to neural signal integrity, creating a dangerous governance gap characterized by systemic misclassification. This paper argues that cognition is rapidly becoming strategic infrastructure, caught between the United States' market-driven diffusion of neurotechnology and China's state-integrated fusion of AI and brain science. Using Singapore as a critical stress test, and applying institutional classification analysis and regulatory mandate mapping, this paper reveals a structural paradox: a state with high regulatory capacity in both cyber and biomedical domains nonetheless remains vulnerable at their intersection due to a failure to classify the human mind as infrastructure. We introduce the concept of *cognitive sovereignty*—the strategic capacity to protect neural processes from external modulation—and propose a *cognitive operational technology* (OT) framework to secure the human mind as a distinct layer of critical national infrastructure.

# 1. INTRODUCTION

As brain-computer interfaces (BCIs) transition from clinical settings to consumer wearables and military-adjacent technologies, they introduce a security domain for which the modern state is structurally unprepared. The convergence of neuroscience and cybersecurity creates a new vulnerability class in which the human nervous system becomes a networked, contestable substrate. As wireless protocols enable the modulation of cognitive function, the distinction between "hacking a device" and "hacking a mind" evaporates.

This shift occurs within a fragmented governance architecture. As Klein and Patrick observe regarding artificial intelligence, emerging technologies are governed not by a coherent authority but by a loose array of overlapping institutions known as a *regime complex*.[1] In neurotechnology, this complexity creates a "dead zone" where civilian and military applications are virtually indistinguishable.[2] Existing frameworks for data privacy, biomedical safety, and critical infrastructure protect specific components, yet none are designed to protect cognition as a strategic asset.

The urgency of this gap is amplified by divergent global strategies. The United States pursues a vertical, market-driven approach relying on rapid diffusion and *ex post* enforcement.[3] Conversely, China adopts a centralized *ex ante* model by integrating algorithmic governance directly into state security. This bifurcation presents a critical dilemma: Can a technologically sovereign democracy govern direct human-machine cognition without defaulting to frontier deregulation or authoritarian control?

This paper argues that neurosecurity must be treated not as a subfield of bioethics, but as a missing layer of national security. We examine Singapore as a critical "stress test" for democratic neuro-governance due to its high regulatory capacity. The analysis reveals a paradox: while Singapore is prepared for cyber and biomedical risks independently, it remains vulnerable at their intersection. We test the claim that *cognitive sovereignty*, or the protection of neural integrity, falls structurally between the Personal Data Protection Act (PDPA) and the Cybersecurity Act. As Bernardi et al. argue, when capability-modifying interventions become infeasible due to diffusion, society must shift toward building resilience against inevitable risks in the form of *adaptation*.[4] Singapore's current regime protects static biometric data,[5] but fails to account for dynamic neural output, thus leaving the cognitive layer of national infrastructure exposed. Methodologically, this paper employs qualitative institutional analysis of statutory definitions, agency mandates, and regulatory jurisdictional boundaries to identify classification failures at the cyber–biological interface.

The paper proceeds as follows: Section 2 defines the theoretical framework of cognitive sovereignty; Section 3 establishes the US-China comparative baseline; Section 4 analyzes Singapore's structural vulnerabilities; and Sections 5 and 6 propose a Neuroethical Cybersecurity Framework to operationalize the nervous system as critical infrastructure.

## 2. THEORETICAL FRAMEWORK: NEUROSECURITY & COGNITIVE SOVEREIGNTY

The transition of BCI technology from clinical therapy to consumer application precipitates a collapse of distinct regulatory categories. Traditionally, the governance of the human body and the governance of digital networks have operated as separate domains. Medical devices are regulated for safety and efficacy,[6] while digital networks are regulated for integrity and access.[7] However, interconnected neural devices violate this separation by transforming the biological substrate of the brain into a networked node.

**2.1 The Neurosecurity Dead Zone**

This convergence creates a governance "dead zone," mirroring the arms control challenges inherent in dual-use technology where civilian and military applications become virtually indistinguishable.[8] In this zone, the same neural decoding algorithms used to restore speech in a patient can be repurposed for surveillance or interrogation. Consequently, BCIs collapse three distinct risk domains—cybersecurity (network integrity), biomedical safety (biological integrity), and data protection (information privacy)—into a single, unresolved domain of neurosecurity.

Current frameworks are structurally incapable of managing this convergence because they treat risk as residing in the device or the data, rather than the user's agency. As noted in recent scholarship on global AI governance, such technologies are not governed by a coherent authority but by a *regime complex*.[9] Within this complex, the specific threat of adversarial cognition, where a user's neural processes are modulated by external actors, falls through the cracks. A BCI might be medically safe (non-toxic) and cyber-secure (encrypted), yet still allow for the sub-perceptual manipulation of cognitive feedback loops,[10] a vector that no existing regime is mandated to monitor.

**2.2 Defining Cognitive Sovereignty**

We define *cognitive sovereignty* not merely as a human right to mental privacy, but as a strategic imperative: the capacity of a state or individual to protect the integrity of neural processes against unauthorized external modulation.

This definition draws on the theory of *weaponized interdependence*. Just as global financial networks contain "chokepoints" that states can exploit for coercion, the digitization of cognition creates *cognitive chokepoints*: proprietary cloud platforms and BCI interfaces upon which human thought increasingly relies.[11] If the brain becomes a node in a network, it becomes subject to the logic of network power; access to the neural substrate can be weaponized to degrade decision-making capability or to extract intelligence.

Operationally, cognitive sovereignty requires protections across three dimensions:

1. *Legal Control***:** The ownership of neural data not merely as "biometric" identification, but as dynamic cognitive output.
2. *Technical Integrity***:** The assurance that neural signals are neither intercepted nor injected with adversarial noise during transmission.
3. *Political Agency***:** The protection of the cognitive substrate from foreign manipulation, ensuring that decision-making processes remain autonomous from covert external influence.

This framework is not intended as a theory of consciousness or personhood, but as a security-governance taxonomy for adversarial digital-biological systems.

**2.3 Misuse Diffusion and the Adaptation Window**

The urgency of establishing this framework is driven by the rapid diffusion of the underlying technologies. Viewed through the lens of general purpose technologies (GPTs), the strategic impact of neurotechnology will depend less on who invents it and more on its diffusion capacity across the economy.[12] However, widespread diffusion introduces a critical vulnerability:

the "pacing problem." As the cost of developing advanced capabilities decreases, capability-modifying interventions, such as restricting access to powerful models, become structurally infeasible.[13] The proliferation of consumer-grade BCIs means that high-fidelity neural data will soon be generated outside the controlled environments of hospitals or research labs.

This creates a compressed adaptation window. Risks associated with advanced cognitive systems diffuse faster than democratic institutions can debate and enact oversight. Unlike nuclear proliferation, where fissile material is scarce and observable,[14] the software-defined nature of neurotechnology allows vulnerabilities to spread instantly through global supply chains. Society is thus forced into a posture of *adaptation* rather than prevention.[15] If a state waits for a catastrophic neuro-cyber event to legislate, it will have already ceded cognitive sovereignty to the market actors controlling the infrastructure.

**2.4 Regime Fragmentation**

Finally, the pursuit of cognitive sovereignty is complicated by the fragmented nature of the international order. There is no dedicated "neuro-governance" layer in international law.[16] Instead, states must navigate a landscape of asymmetrical understanding, where terms like "regulation" mask vastly different strategic intents.

The current *regime complex* is split between incompatible normative frameworks. The market-driven model prioritizes innovation speed and *ex post* enforcement, hence treating safety as a product feature to be solved by market mechanisms.[17] Conversely, the state-integrated model prioritizes *ex ante* control,[18] viewing cognitive data as a national resource to be managed for stability and industrial dominance.[19] Because the necessary international norms are absent, highly digitized nations are left vulnerable. They cannot depend on existing agreements to shield their citizens from cognitive warfare or exploitation by corporations, leaving them in a precarious position due to this regulatory fragmentation.

This framework establishes the conceptual tools used in the analysis; Section 5 applies these tools to diagnose how the neurosecurity governance gap manifests empirically in Singapore.

## 3. THE TWO GLOBAL EXTREMES: MARKET DIFFUSION VS. STRATEGIC INTEGRATION

To understand the structural vulnerability of a small, highly digitized state like Singapore, one must first map the diverging neuro-strategic pressures radiating from the two artificial intelligence superpowers. The global neurotechnology landscape is currently defined by a schism between the United States' market-driven diffusion and China's statist integration.[20] This divergence creates a dynamic wherein the global supply chains for cognitive infrastructure become mechanisms of coercion rather than mere trade channels, following the logic of weaponized interdependence.[21] Access to the neural substrate is rapidly becoming a strategic chokepoint, forcing third-party states to navigate between the chaos of the frontier market and the control of the surveillance state.

### 3.1 United States: Frontier Market Neurosecurity

The United States pursues a strategy defined by "frontier market neurosecurity," treating BCIs as a nascent GPT.[22] Where nuclear weapons required containment, the U.S. strategy for GPTs, as Jeffrey Ding argues, relies on *diffusion capacity*.[23] However, this reliance on diffusion creates a specific vulnerability: the "pacing problem."[24] As aforementioned in Section 2.3, such conditions minimize the feasibility of capability-modifying interventions, such as safety regulations.[25]

The U.S. regulatory landscape remains stubbornly vertical. While export controls through the Bureau of Industry and Security (BIS) increasingly constrain the international diffusion of advanced neuro-hardware[26] and the FDA has expanded cybersecurity review authority for medical devices,[27] these measures do not resolve the structural fragmentation of domestic neurodata governance. This fragmentation leaves the "cognitive data" produced by consumer BCIs in a legal gray zone, protected neither by HIPAA, which applies only to covered medical entities,[28] nor by comprehensive federal privacy statutes. The result is a form of frontier chaos where cognitive agency is subordinated to the cultural "social fact" of market efficiency.[29] By treating safety as a product feature to be solved by market mechanisms, the U.S. model creates systemic functional risks in the form of degraded user agency through unmonitored commercial feedback loops.

### 3.2 China: State-Integrated AI-Brain Control

In contrast, China is constructing a regime of "state-integrated AI–brain fusion."[30] This iterative regulatory project views cognition as a domain of "industrial, informational, and military power," employing strategies of "preemptive security" to render the functional parameters of AI legible to the state.[31] As a result, China's governance has evolved from broad mandates to highly specific regulations targeting recommendation algorithms and "deep synthesis" technologies.[32] This integration facilitates a "civil-military fusion" that effectively erases the distinction between a medical BCI and a tool for cognitive monitoring.[33]

For a third-party state, China's model presents the threat of technological lock-in. If the brain becomes a node in a global network, adopting Chinese neuro-infrastructure, whether EEG headsets or neuromorphic chips, risks embedding foreign content controls or data export protocols directly into the firmware of the device.[34] This is not an abstract concern about "truthfulness" but a technical risk regarding hard-coded signal processing parameters that filter or prioritize neural data before it ever reaches the user. Singapore faces a structural crisis in this bifurcation: high adoption capacity, usually a strategic asset,[35] here accelerates the speed at which its population becomes exposed to either the extractive data practices of the U.S. market or the centralized data aggregation of the Chinese state.

## 4. SINGAPORE AS THE NEUROSECURITY STRESS TEST

If the United States represents the risks of market diffusion and China represents the rigors of state integration, the Republic of Singapore offers a critical third case: the "smart nation" operating at maximum regulatory capacity.[36] Singapore functions as the ideal empirical stress test for the concept of neurosecurity. This section employs qualitative institutional analysis of Singapore's cybersecurity, biomedical, and data protection regimes. Regulatory mandates, statutory definitions, and agency jurisdictional boundaries are analyzed to identify classification failures at the cyber-biological interface. Despite possessing a centralized cybersecurity architecture (CSA) and a world-class biomedical hub (Biopolis), this analysis reveals that even such high capacity is insufficient to secure cognitive sovereignty without a specific doctrinal shift toward protecting the mind as infrastructure.

The core of Singapore's vulnerability lies in its neuro-biomedical ecosystem, anchored in the Biopolis and Fusionopolis[37] complexes. Here, the Agency for Science, Technology and

Research (A*STAR) actively develops BCIs for clinical and assistive applications under the "Research, Innovation and Enterprise 2025" (RIE2025) plan.[38] While robust, this ecosystem is structurally exposed via the transmission mechanism of global supply chains. Singapore's strategy positions it as a trusted node in the global flow of biomedical data,[39] but as Abraham Newman notes regarding interdependent networks, this connectivity creates strategic vulnerabilities.[40] By relying on foreign-designed BCI chips or cloud processing platforms, Singapore's burgeoning neuro-sector inadvertently imports the security flaws of its suppliers and threatens the cognitive sovereignty of its population.

The state's defense against digital threats is formidable but categorically misaligned with this new threat. The Cyber Security Agency (CSA) operates under a comprehensive mandate established by the *Cybersecurity Act of 2018*, designating eleven sectors as Critical Information Infrastructure.[41] This framework reflects high adoption capacity,[42] but the CSA's mandate is structurally limited to the protection of networks and physical systems. The *Operational Technology (OT) Cybersecurity Masterplan 2024* explicitly defines its scope as systems that monitor and control physical processes in order to prioritize the availability and integrity of industrial hardware.[43] By defining critical infrastructure strictly in terms of physical plant operations, the Masterplan inadvertently excludes the *biological* processors (human operators) that interface with these systems. Consequently, a BCI is treated merely as a peripheral edge device rather than a gateway to a critical asset, leaving the "cognitive layer" outside the perimeter of national defense.

This gap is further widened by the structural limitations of the *Personal Data Protection Act* (PDPA). While often lauded for its balance, the PDPA falls into a "biometric trap" when applied to neurosecurity. The Personal Data Protection Commission (PDPC) guidelines explicitly confine biometric samples to "physiological, biological, or behavioural characteristics of an individual."[44] This definition creates a legal blind spot: it protects neural data only when used for *authentication* (identity), not when used for *actuation* (agency). Because the law focuses on preventing identity theft rather than cognitive sovereignty, it offers no recourse for "sub-perceptual" adversarial attacks that alter a user's decision-making without ever technically breaching data confidentiality or misidentifying the user.

The Health Sciences Authority (HSA) regulates medical devices with rigor comparable to the U.S. FDA,[45] yet it too misses the specific nature of the threat. The HSA's *Regulatory Guidelines for Telehealth Products* mandate cybersecurity measures to prevent "unauthorized access" that could compromise "patient safety."[46] However, the guidelines interpret safety through a clinical lens (i.e. preventing physical injury, toxicity, or misdiagnosis), rather than a cognitive lens. This framework asks whether a hack will stop a pacemaker from beating, but fails to ask whether a hack will change a user's intent. This "functional integrity" gap means a device can be fully compliant with HSA safety regulations while failing to inhibit cognitive warfare.

Ultimately, the Singapore case demonstrates a failure to implement cognitive operational technology (OT). If the human mind in a BCI loop functions as a node in a network, then the nervous system must be treated as the ultimate form of OT that controls the physical reality of the state through the actions of its citizens. Currently, incentive structures in Singapore's startup ecosystem prioritize speed-to-market,[47] treating security as a compliance checkbox rather than a strategic necessity. Startups like Neeuro[48] or SynPhNe[49] operate outside the heavy regulatory burden of Critical Information Infrastructure via commercialization. Without a doctrine of cognitive OT to govern these interfaces, Singapore, although having built robust frameworks for cyber and biological risks, paradoxically remains structurally unprepared for the intersection of the two.

## 5. THE NEUROSECURITY GOVERNANCE GAP

The Singapore case demonstrates that the neurosecurity governance gap is produced by systematic misclassification of the object being governed rather than regulatory incompetence. Despite possessing high state capacity, Singapore's regulatory architecture fractures because each agency governs a constituent component of the BCI system while missing the emergent property of the system itself: cognitive agency. This results in a taxonomy failure where the specific mechanics of adversarial cognition remain invisible to security planners because they do not fit the existing definitions of harm.

### 5.1 The Data Misclassification: From Identity to Actuation

The Personal Data Protection Act (PDPA) governs the informational record. Its mandate is to ensure privacy and consent, measuring success by the prevention of unauthorized data disclosure.[50] Structurally, however, it does not govern cognitive actuation. By categorizing neural data as "biometrics," the regime misclassifies dynamic thought processes as static identity markers. This legal framework is designed to prevent a malicious actor from stealing a user's fingerprint to unlock a phone; it is not designed to prevent a malicious actor from using a feedback loop to subtly alter the user's risk tolerance or emotional state. Consequently, the PDPA protects the *file* while leaving the *mind* open to modification, and creates a vulnerability where cognitive manipulation is legally permissible as long as the data remains encrypted and "private."

### 5.2 The Biomedical Misclassification: From Safety to Integrity

The Health Sciences Authority (HSA) governs the physical device. Its mandate is clinical safety, measuring success by the absence of toxicity, thermal injury, or electrical malfunction.[51] Structurally, it does not govern adversarial intent. A neuro-device can be medically safe, causing no physical harm to the brain tissue, while simultaneously functioning as a vector for cognitive warfare. By viewing the brain solely as a biological organ rather than a decision-making node, the HSA misclassifies neurosecurity risks as "physiological" rather than strategic. This creates a blind spot where a compromised BCI that biases a user's decision-making without causing physical pain is treated as a compliant medical device rather than a weaponized instrument.

### 5.3 The Infrastructure Misclassification: From Network to Node

The Cyber Security Agency (CSA) governs the digital network. Its mandate is the integrity and availability of Critical Information Infrastructure, measuring success by network uptime and resistance to penetration.[52] Structurally, it does not govern the human operator as a component of that infrastructure. By defining OT strictly in terms of physical plants and servers, the CSA misclassifies the human mind as "non-infrastructure." Once again, this leaves the ultimate decision-making processor, the human user interfacing with the grid via a BCI, outside the security perimeter. The regime protects the switch, yet ignores the mind controlling the hand that flips it.

Defenders of the existing framework might argue that the combination of PDPA, HSA, and CSA mandates creates a comprehensive safety net that jointly covers the full lifecycle of BCI risk. This perspective erroneously assumes that privacy, patient safety, and network security are separable domains, as real-time cognitive actuation proves otherwise. In a closed-loop BCI, a privacy breach (data extraction) can be functionally identical to a safety breach (unauthorized modulation) and a security breach (network penetration). By treating these as distinct regulatory silos, the current framework fails to address the emergent property of the system: the continuous, real-time integration of digital instructions with biological execution.

**5.4 The Normative Vacuum**

This triple misclassification, treating agency as identity, manipulation as medicine, and minds as users, sustains the governance gap. It persists because, as Finnemore and Sikkink argue regarding the lifecycle of international norms, global governance is reactive.[53] Currently, there is no international norm of "cognitive non-interference" to force a domestic reclassification. This normative vacuum allows states and corporations to occupy the "pre-AGI zone" of human cognition without violating formal international law.[54] In the absence of a clear norm defining where the "human" ends and the "machine" begins, the nervous system becomes a gray zone for hybrid warfare. Consequently, in this domain, an adversary can degrade a population's capacity to reason without ever triggering a conventional act of war.

**6. THE NEUROETHICAL CYBERSECURITY FRAMEWORK**

If the current global governance architecture is defined by a "dead zone" regarding dual-use technologies, as Vaynman and Volpe suggest,[55] then the solution cannot be a mere expansion of existing privacy laws or medical safety standards. Instead, nations must construct a specific Neuroethical Cybersecurity Framework to bridge the gap between biomedical safety and national security. This framework operates on the premise of "societal adaptation," a concept Bernardi et al. apply to advanced AI, arguing that when technological diffusion outpaces the feasibility of prevention, governance must shift toward resilience mechanisms.[56]

The first line of defense is technically focused on the enforcement of distinguishability. Considering stability in military AI requires mechanisms to distinguish between benign and lethal systems,[57] neuro-devices need to be designed to clearly differentiate between actions that provide therapeutic benefits, like stopping a tremor, and those that modulate cognition, such as changing a user's mood. To achieve this, BCI data streams require homomorphic encryption to enable secure processing alongside "neural watermarking" to verify biological signal origin. It is important to note that these mechanisms are not proposed as near-term compliance requirements for consumer BCIs, but as forward-architecture constraints to prevent irreversible insecurity as hardware latency and compute ceilings advance. Establishing these benchmarks now ensures that "security by design"[58] architecture is in place to prevent the "man-in-the-brain" attacks identified in the governance gap.

Legally, democracies like Singapore must move beyond the "biometric trap" inherent in current data protection regimes. To address the unique risks of direct neural interfacing, it is optimal for legislation to codify "neurodata" as a distinct protected class that is separate from biological or biographical data. This distinction necessitates a new standard of coercion-resistant consent. Current "click-wrap" consent models are insufficient for devices that access the biological substrate of consent itself. Therefore, legal standards must mandate continuous, revocable consent protocols capable of detecting if a user's capacity to opt-out has been neurally degraded. Following the advocacy of the "neurorights" movement,[59] legislation must explicitly criminalize unauthorized neural interference, interpreting such a definition not merely as a violation of privacy, but rather as a fundamental attack against the individual.

Institutionally, the framework requires operationalizing the concept of cognitive operational technology. To operationalize cognitive sovereignty, states must treat the neuro-sphere with the same gravity as critical physical infrastructure. This begins with pre-market neurosecurity audits. Just as the FDA mandates clinical trials for safety,[60] a dedicated cognitive security-oriented administration must mandate adversarial testing to identify capabilities that could be exploited for cognitive warfare.[61] Furthermore, this institutional oversight requires the creation of a specialized Computer Emergency Response Team (CERT). Existing CERTs focus on network uptime and data integrity;[62] a dedicated neuro-CERT unit is required to monitor for patterns of population-level neural affect that might suggest a coordinated cognitive warfare campaign.

Finally, given the persistent deadlock between the American and Chinese regulatory models, a universal treaty remains unlikely. Instead, it is most feasible and effective for small, highly developed states to pursue *minilateral arrangements*.[63] Additionally, a coalition of neuro-sovereign nations could form what Vaynman and Volpe call an *AI Cartel*[64] for neurotechnology by setting high import/export standards. This forces manufacturers to adopt safety protocols to access profitable markets.

Regulatory states that are small yet highly effective wield a significant impact. These entities exercise softer power via setting fundamental benchmarks and operational guidelines. If Singapore aligns neurosecurity standards with similarly positioned jurisdictions such as Switzerland and the European Union, manufacturers are likely to comply since maintaining divergent firmware, certification, and firmware-security pipelines for different markets is commercially non-viable at scale.

## 7. IMPLICATIONS: WHAT SINGAPORE TEACHES THE WORLD

The case of Singapore clarifies the stakes for the rest of the world. Singapore possesses a world-class cybersecurity agency (CSA), a rigorous biomedical regulator (HSA), and a comprehensive data protection regime (PDPA). Yet, as demonstrated, these separate pillars create a significant pre-deployment security externality by leaving the "cognitive gate" unguarded. If Singapore, with its high state capacity, struggles to close the neurosecurity gap, this structural vulnerability is acute for larger, less agile democracies.

The Singaporean experience warns against the complacency of vertical regulation, demonstrating that adoption capacity does not equate to resilience. Rapidly integrating foreign neurotechnology without a corresponding cognitive OT doctrine accelerates the erosion of sovereignty and exposes the state to an anticipatory threat class of cognitive interference. Crucially, this threat is distinct from psychological operations (PsyOps); cognitive warfare via BCIs involves the direct, technical modulation of neural function.

However, Singapore also offers a solution. By pivoting its existing infrastructure toward neuro-resilience through integrating the CSA and HSA, Singapore can pioneer a "third way" between American frontier chaos and Chinese state control. This strategic pivot involves transcending traditional ministerial silos by reclassifying the cognitive substrate as a critical domain of national infrastructure. Such a move would secure Singapore's cognitive sovereignty against modeled adversarial risks and provide a replicable model for other states navigating the polarized global technology landscape.

## 8. CONCLUSION

As BCIs transform the human nervous system into a domain of national security, this paper has demonstrated that the convergence of AI and neuroscience exposes a systemic misclassification within global governance, where the specific risks of adversarial cognition fall between the cracks of cyber, medical, and data law. To address this, we introduced *cognitive sovereignty*: a strategic posture that moves beyond device security to protect the capacity for independent thought against the structural vulnerabilities introduced by global supply chains. However, this reclassification is not without peril; treating the mind as critical infrastructure introduces residual risks of state abuse through excessive monitoring, corporate capture of regulatory standards, and complex civil liberties trade-offs between collective security and mental privacy. In this "pre-AGI zone" of high ambiguity, the failure of democracies to navigate these tensions will cede the norms of cognitive liberty to the market efficiencies of Silicon Valley or the surveillance imperatives of Beijing. Ultimately, preserving human autonomy requires establishing robust cognitive operational technology frameworks that secure the neural-digital interface as a domain of agency rather than a vector of control. The next frontier of human rights is the mind itself, and protecting it is the essential task of our generation.

# Endnotes